\documentclass[aps,showpacs,twocolumn,superscriptaddress]{revtex4}
\usepackage[dvips]{graphicx}
\usepackage{amsmath}
\usepackage{color}

\begin{document}

\title{Energy carriers in the Fermi-Pasta-Ulam $\beta$ lattice: Solitons or Phonons?}

\author{N. Li}
\affiliation{Max Planck Institute for the Physics of Complex
Systems, N\"othnitzer Strasse 38, D-01187 Dresden, Germany}

\author{B. Li}
\affiliation{NUS Graduate School for Integrative Sciences and
Engineering, Singapore 117456, Republic of Singapore}
\affiliation{Department of Physics and Centre for Computational
Science and Engineering, National University of Singapore, Singapore
117546, Republic of Singapore}

\author{S. Flach}
\affiliation{Max Planck Institute for the Physics of Complex
Systems, N\"othnitzer Strasse 38, D-01187 Dresden, Germany}

\begin{abstract}
We investigate anomalous energy transport processes in the
Fermi-Pasta-Ulam $\beta$ lattice. They are determined by the maximum
sound velocity of the relevant weakly damped energy carriers. That
velocity can be numerically resolved by measuring the propagating
fronts of the correlation functions of energy/momentum fluctuations
at different times. The numerical results are compared with the
predictions for solitons and effective (renormalized) phonons,
respectively. Excellent agreement has been found for the prediction
of effective long wavelength phonons, giving strong evidence that
the energy carriers should be effective phonons rather than
solitons.
\end{abstract}
\date{\today}

\maketitle

Energy transport in low dimensional systems has attracted enduring
interest \cite{ht1,ht2,ht3,ht4}. One striking finding is the
phenomenon of anomalous transport \cite{at1,at2,at3}, which has
recently been experimentally verified for carbon nanotubes
\cite{CWChang1}. Theoretical efforts \cite{ht2,te1,te2} usually
follow the pioneering work of Peierls and focus on the low
temperature region, where weakly interacting phonons are considered
to be the responsible energy carriers. A microscopic transport
theory beyond the low temperature regime is still lacking, which
leaves the explanation for most existing numerical and experimental
results far from satisfactory. One central question concerns the
type of energy carriers whose properties determine the underlying
transport behavior at higher temperatures. As temperature or
nonlinearity is increased, collective motions other than phonons
could also be excited. It is thus desirable to identify the specific
energy carriers for these low dimensional systems.

The Fermi-Pasta-Ulam $\beta$ (FPU-$\beta$) lattice is a classic
example showing the effect of anomalous transport, and therefore a
perfect testbed for comparison between theoretical predictions and
numerical experiments. Anomalous transport manifests through an
increase of the heat conductivity with the system size. This in turn
implies that the responsible energy carriers are anomalously weakly
damped and propagate ``ballistically" over very long distances. Due
to nonlinearity,
solitons \cite{friesecke,st1}, discrete breathers
\cite{bt1} and interacting phonons
\cite{ept1,ept2,ept3,lnb1,ept4}  are candidates for these carriers.
In particular, supersonic
solitons have been considered as major energy carriers which are
responsible for the anomalous transport behavior
\cite{soliton1,soliton2}. The pivotal evidence supporting the idea
of soliton transport is the numerical observation of ultrasonic
energy transfer. The sound velocity $c_s$ of energy transfer was
measured by following the spreading of an initial energy pulse,
using both non-equilibrium \cite{soliton1,bwli1} and equilibrium
\cite{soliton2} methods. In particular, the temperature-dependent
sound velocity $c_s$ is compared with a prediction derived from
soliton theory \cite{st1} and good agreement has been found in Ref.
\cite{soliton1}. However, strong finite size effects of the soliton
velocities were not clarified in Ref. \cite{st1}.
The same
data for $c_s$ in Ref. \cite{soliton1} are also in good
agreement with the predicted velocity for effective phonons
\cite{lnb1}.
The uncertainty of the computed data in Ref. \cite{soliton1} is too
large to distinguish between the two predictions. To identify the
true energy carriers, a more accurate numerical determination of
$c_s$ is needed.

In the present paper, we apply the equilibrium approach recently
developed by Zhao \cite{soliton2} to study the energy transport
properties in the FPU-$\beta$ lattice. As demonstrated in Ref.
\cite{soliton2}, the sound velocities $c_s$ can be measured with
very high precision. We will show that the numerical results are in
very good agreement with the prediction for effective phonons and
the agreement is not limited to the FPU-$\beta$ lattice. The soliton
predictions show clear deviations and can be ruled out.

We consider the dimensionless Hamiltonian for the
FPU-$\beta$ lattice
\begin{equation}
H=\sum^{N}_{i=1}\left[\frac{p^2_i}{2}+\frac{1}{2}(q_i-q_{i-1})^2+\frac{1}{4}(q_i-q_{i-1})^4\right]
\end{equation}
where $p_i$ denotes the momentum and $q_i$ denotes the displacement
from equilibrium position for the $i$-th atom with
$i=0,\pm1,\pm2,...,\pm N/2$. The local energy density is defined as
$H_i=p^2_i/2+(q_i-q_{i-1})^2/2+(q_i-q_{i-1})^4/4$. Fixed boundary
conditions are applied to the two end atoms, which are additionally
coupled to stochastic Langevin heat baths with specified temperature
$T$. The normalized correlation functions of energy and momentum
fluctuations are defined as \cite{soliton2}
\begin{equation}
C_E(i,t)=\frac{\left<\Delta H_i(t)\Delta H_0(0)\right>}{\left<\Delta
H_0(0)\Delta
H_0(0)\right>},C_P(i,t)=\frac{\left<p_i(t)p_0(0)\right>}{\left<p_0(0)p_0(0)\right>}
\end{equation}
where $\Delta H_i(t)\equiv H_i(t)-\left<H_i\right>$, and
$\left<\cdot\right>$ denotes ensemble averages. Note that
$C_{E/P}(i,t=0)=\delta_{i,0}$.
Therefore, the correlation functions $C_{E/P}(i,t)$ describe
the spatiotemporal spreading of the initial energy/momentum
fluctuations \cite{soliton2}.

\begin{figure}
\centerline{\includegraphics[width=8cm]{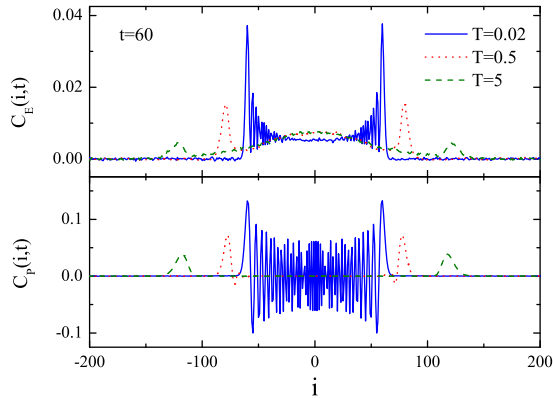}} \caption{(color
online) Spatial distributions of the correlation functions of
$C_{E}(i,t)$ (upper panel) and $C_{P}(i,t)$ (lower panel) at $t=60$
for $T=0.02,0.5$ and $5$ for the FPU-$\beta$ lattice with $N=500$.}
\label{fig1}
\end{figure}

In Fig. \ref{fig1}, we depict the spatial distributions of the
correlation functions $C_{E/P}(i,t)$ at $t=60$ for different temperatures. The values $T=0.02,0.5$ and $5$
correspond to
low, intermediate and high temperature regimes, according to
the
scaling behavior of the aspect ratio
$\epsilon=\left<(q_i-q_{i-1})^4\right>/\left<(q_i-q_{i-1})^2\right>$.
The aspect ratio scales as $\epsilon\propto T$ in the low temperature
regime and as $\epsilon\propto T^{1/2}$ in the high temperature regime.
As can be seen from Fig. \ref{fig1}, both distributions
possess symmetric propagating fronts with identical propagation velocities.
These propagating fronts are induced by the fastest travelling energy carriers
\cite{soliton1,soliton2}.
With increasing temperature the propagation velocity increases,
which is caused entirely by the presence of nonlinear terms in the equations of motion.
The
fluctuations of $C_{P}(i,t)$ are much smaller than those of
$C_{E}(i,t)$. Therefore, we will determine the sound velocity $c_s$
by measuring the peak positions of the propagating fronts for $C_{P}(i,t)$, as
in Ref. \cite{soliton2}. We have also tested that the results
do not depend on the type of boundary conditions (fixed or periodic).

\begin{figure}
\centerline{\includegraphics[width=8cm]{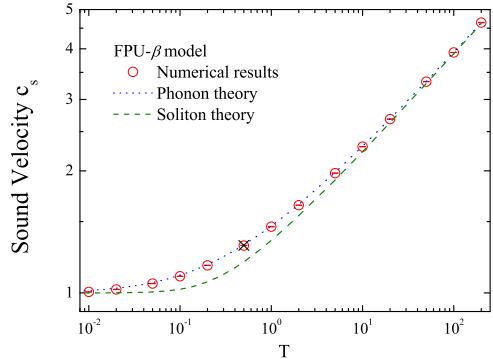}} \caption{(color
online) Sound velocity $c_s$ as the function of temperature $T$ for
the FPU-$\beta$ lattice with $N=1000$. The circles are the numerical
results. Error bars have been plotted, but they are much smaller
than the symbol size. The dotted line is the analytic prediction for
effective phonons from Eq. (\ref{ph}) and the dashed line is the
analytic prediction for solitons from Eq. (\ref{sl}) with
$\eta=2.215$. The numerical result obtained for $T=0.5$ in Ref.
\cite{soliton2} is plotted with a cross symbol.} \label{fig2}
\end{figure}

Let us briefly introduce the predictions of sound velocity for
solitons and effective phonons. According to \cite{friesecke,st1},
the soliton profile with $z \sim q_i-q_{i-1}$ is given by
\begin{equation}
 Q_s(z) = \sqrt{2(c_s^2-1)}{\rm sech} \left(2z\sqrt{(c_s^2-1)/c_s^2}\right)\;.
\end{equation}
It follows that the energy of a soliton is proportional to
$c_s^3\sqrt{c_s^2-1}$, where we assume $c_s>0$ without loss of
generality. Using a Boltzmann distribution for energies of
excitations, we conclude that the temperature $T$ introduces an
energy scale such that larger soliton energies are exponentially
suppressed. At the same time smaller energies imply smaller sound
velocities. Therefore a Boltzmann distributed gas of solitons will
typically show maximum sound velocities which correspond to solitons
with an energy of the order of the temperature $T$. Therefore it
follows that
%
\cite{soliton1}
\begin{equation}\label{sl}
c^3_s\sqrt{c^2_s-1}=\eta T\;.
\end{equation}
The constant $\alpha$ is a free fit parameter.
Note that $c_s\rightarrow 1$ as $T\rightarrow 0$
and $c_s\approx \eta^{1/4} T^{1/4}$ as $T\gg 1$ in the high
temperature regime.

For the effective phonons, the sound velocity is defined as the
maximum group velocity of the renormalized phonons. Here,
renormalization implies a mean field treatment of nonlinear terms in
the equations of motion. As a result eigenfrequencies of phonons are
renormalized, and will increase with increasing temperature.
Therefore, renormalized phonons will also yield sound velocities
which increase with increasing temperature, becoming supersonic as
compared to the case of $T\rightarrow 0$. In particular,
$c_s=\partial{\hat{\omega}_k}/\partial{k}|_{k=0}$ where
$\hat{\omega}_k=2\sqrt{\alpha}\sin{k/2}$ with $0\leq k<2\pi$ and
$\alpha=1+\left<\sum_i (q_i-q_{i-1})^4\right>/\left<\sum_i
(q_i-q_{i-1})^2\right>$
\cite{lnb1}. It follows \cite{lnb1}
\begin{equation}\label{ph}
c_s=\left(1+\frac{\int^{\infty}_0 x^4
e^{-(x^2/2+x^4/4)/T}dx}{\int^{\infty}_0 x^2
e^{-(x^2/2+x^4/4)/T}dx}\right)^{\frac{1}{2}}\;.
\end{equation}
For $T \rightarrow 0$ we find $c_s\rightarrow 1$, and in the high
temperature region $c_s\approx \sqrt{\int^{\infty}_0 x^4
e^{-x^4/4T}dx/\int^{\infty}_0 x^2 e^{-x^4/4T}dx}\approx 1.22
T^{1/4}$.
Both predictions (solitons and phonons) yield three
similar results for the sound velocity: (i) $c_s \geq 1$;
(ii) $c_s(T\rightarrow 0) \rightarrow 1$;
(iii) in the high temperature regime, the sound velocities exhibit the
same scaling with temperature as $c_s\propto T^{1/4}$.

In Fig. \ref{fig2}, we plot the numerically determined sound
velocity $c_s$ as the function of temperature $T$ for the
FPU-$\beta$ lattice. The computational errors are extremely small
and $c_s$ is measured very accurately. The numerical results are
compared with the predictions for solitons from Eq. (\ref{sl}), and
for effective phonons from Eq. (\ref{ph}). Excellent agreement has
been observed for the effective phonon result in the entire
temperature region being explored. The soliton curve corresponds to
$\eta=2.215$, which reproduces the correct high temperature result.
However,
the deviation from the prediction for solitons of Eq. (\ref{sl}) is
quite distinct in the intermediate temperature regime where
$0.05\leq T\leq 2$. We note that variations and optimizations of
$\eta$ do not improve this discrepancy. The cross symbol in Fig.
\ref{fig2} represents the sound velocity $c_s=1.31$ measured at
$T=0.5$ in Ref. \cite{soliton2}. Although it was originally viewed
as an evidence for soliton transport, the numerical result is
actually coinciding both with our numerical data and with the
effective phonon results, in contrast to the soliton theory.
This finding
provides strong evidence that the effective phonons, rather than the
solitons, should be the energy carriers responsible for anomalous transport in the FPU-$\beta$ lattice.

\begin{figure}
\centerline{\includegraphics[width=8cm]{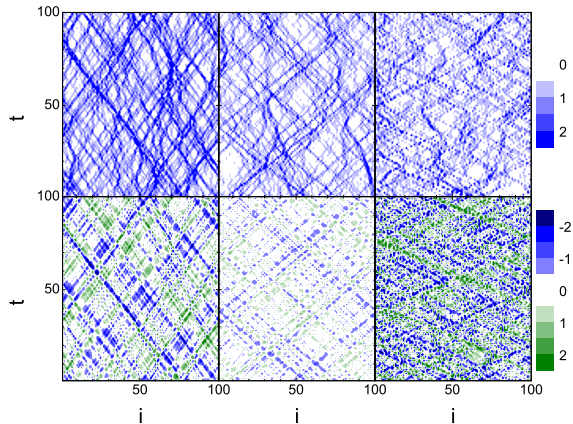}} \caption{(color
online) Spatiotemporal evolution of energy densities $H_i(t)/T$
(upper panels) and relative displacements $q_i(t)-q_{i-1}(t)$ (lower
panels) at thermal equilibrium. The left, middle and right columns
correspond to the harmonic lattice at $T=1$, and the FPU-$\beta$
lattice at $T=1$ and $T=20$, respectively. The lattice size $N=100$
and periodic boundary conditions are applied.} \label{fig3}
\end{figure}

To visualize the energy transport processes, in Fig. \ref{fig3} we
plot the spatiotemporal evolutions of local energy densities
$H_i(t)/T$ and the relative displacement $q_i(t)-q_{i-1}(t)$ for the
harmonic lattice and the FPU-$\beta$ lattice, respectively. The
systems are thermalized at a given temperature and then the heat
bath is removed. The evolution functions are recorded at thermal
equilibrium and results for a suitable time window are displayed.
The spatiotemporal patterns for both the FPU-$\beta$ lattice and the
harmonic lattice are qualitatively similar. Since  solitons are
definitely excluded for the harmonic lattice, there appears to be no
signature for soliton transport in the FPU-$\beta$ lattice as well.
The qualitative behavior of the spatiotemporal evolutions does not
change for larger time windows.

\begin{figure}
\centerline{\includegraphics[width=8cm]{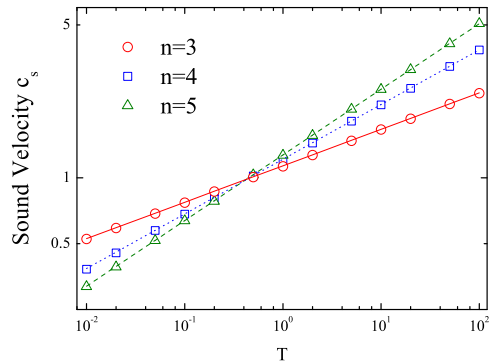}} \caption{(color
online) Sound velocity $c_s$ as the function of temperature $T$ for
the $H_n$ models with $n=3,4$ and $5$. The symbols correspond to the
numerical results at $N=1000$, whereas the lines are the predictions
for effective phonons from Eq. (\ref{v345}). The errors are much
smaller than the symbols.} \label{fig4}
\end{figure}

In the high temperature limit, the FPU-$\beta$ model can be
reduced to an $H_n$ model with $n=4$ assuming the following
Hamiltonian
\begin{equation}
H_n=\sum^{N}_{i=1}\left[p^2_i/2+\left|q_i-q_{i-1}\right|^n/n\right]\;.
\end{equation}
To demonstrate the power and consistency of the effective phonon
formulation, we consider three different cases with $n=3,4$ and $5$.
Following the same procedure as for FPU-$\beta$ lattice, the sound
velocities of effective phonons can be expressed with a compact
formula
\begin{equation}\label{v345}
c_s=\left[\Gamma((n+1)/n)/\Gamma(3/n)\right]^{1/2} (nT)^{1/2-1/n}
\end{equation}
These predictions are plotted in Fig. \ref{fig4} and compared with
numerical results. We again find quantitative agreement for all
three models.

Let us discuss some details in the spatiotemporal dependence of
the correlation functions.
In Fig.
\ref{fig1}, the distribution functions of $C_{E/P}(i,t)$ show
many peaks between the two propagating fronts at $T=0.02$. These
peaks are typical for coherent phonon propagation in harmonic lattices.
For $T=0.5$ and $5$, there are
no visible additional peaks for both distributions and a big hump
emerges in the interior region for $C_{E}(i,t)$.
The disappearance of the intermediate peaks can be therefore attributed
to relaxation processes.
\begin{figure}
\centerline{\includegraphics[width=8cm]{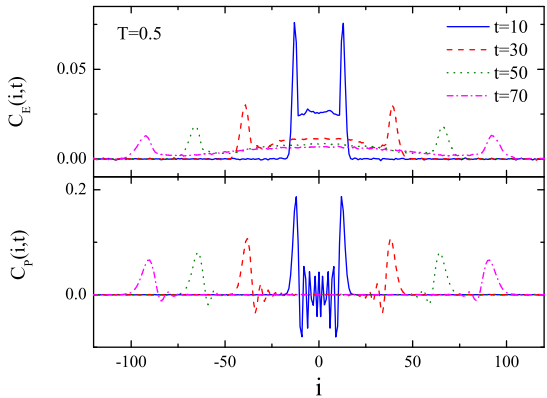}} \caption{(color
online) Spatial distributions of the correlation functions of
$C_{E}(i,t)$ (upper panel) and $C_{P}(i,t)$ (lower panel) at $T=0.5$
for various values of time $t=10,30,50$ and $70$, for the
FPU-$\beta$ lattice with size $N=500$.} \label{fig5}
\end{figure}
Indeed,
the distribution functions $C_{E/P}(i,t)$ for $T=0.5$ at various
spreading times $t=10,30,50$ and $70$ in Fig. \ref{fig5} show that
the short time behavior at $T=0.5$ is similar to that for $T=0.02$.
The phonon modes relax faster at higher temperatures, while on the
other hand, the long wave-length phonon modes possess very long
correlation times even at very high temperatures. According to Ref.
\cite{soliton2}, the correlation function of energy fluctuations
$C_{E}(i,t)$ is nothing but the energy density probability
distribution function (PDF). Therefore we can study the energy
diffusion process
by measuring the mean square displacement (MSD) as
$\left<r^2(t)\right>=\sum_i i^2 C_E(i,t)$. For PDFs shown in Fig.
\ref{fig5}, the MSD is dominated by the area around the propagating
fronts and can be approximated as $\left<r^2(t)\right>\propto
t^{2-\nu}$ where the exponent $\nu$ characterizes the deminishing of
the peak area with time as a consequence of slow but unavoidable
dephasing of even long wave length pohonons.
According to \cite{soliton2} the exponent $0<\nu<1$, and therefore
the diffusion process is
superdiffusive as $\left<r^2(t)\right>\propto t^{\sigma}$ with
$\sigma=2-\nu$. It
is interesting to analyze the connection between superdiffusion and anomalous heat
conduction for the FPU-$\beta$ lattice. Recall that the
heat flux of phonons is $J_k=v_k E_k$ \cite{ht2} where $k$
denotes the wave number. The correlation function of the total heat
flux ($J=\sum_k J_k$) can be approximately obtained as
$\left<J(t)J(0)\right>\approx\left<c^2_s E_0(t)E_0(0)\right>\propto
t^{-\nu}$ since the only long time correlation is due to the
energy carried by the long wave-length phonon modes. Applying the
Green-Kubo formula for heat conductivity \cite{ht2}, we obtain
$\kappa\propto \int^{N/c_s}_0 \left<J(t)J(0)\right> dt \propto
N^{\beta}$ where $\beta=1-\nu$. Without knowing the exact value of
$\beta$ and $\sigma$, we obtain that the  superdiffusion and anomalous heat
conduction are connected via the exponent relation $\beta=\sigma-1$
\cite{soliton2,denisov}.

In conclusion, we have investigated the energy transport processes
for the FPU-$\beta$ lattice using an equilibrium approach. To identify
the energy carriers, we accurately measure the sound velocity of the
energy carriers by following the correlated spreading of the initial
energy/momentum fluctuations. The sound velocities are found to be in
excellent agreement with theoretical predictions for effective
phonons. This predicability has been further confirmed for a series
of $H_n$ models. On the other hand, no signature of soliton
transport has been detected by visualizing the spatiotemporal
evolutions of local energy densities and relative displacements.
Therefore our numerical results clearly reveal that the energy
carriers are long wavelength phonons for the FPU-$\beta$ lattice.

We thank J. D. Bodyfelt, Ch. Skokos, D. O. Krimer and T. Lapteva for
useful discussions.

\end{document}